\documentclass[prb,onecolumn,showpacs]{revtex4}

\usepackage{latexsym,longtable}
\usepackage{amsthm, amsmath, amstext, amsgen, amsbsy, amsopn, amsfonts, amssymb, amsmath, revsymb}
\usepackage{epsfig,graphics,subfigure}

\newcommand{\kp}{\ensuremath{\mathbf{k \cdot p}}}		

\newcommand{\deltaFunction}{\ensuremath{\delta}~function}

\newcommand{\eqRef}{Eq.~}

\newcommand{\figRef}{Fig.~}

\newcommand{\unit}[1]{\ensuremath{\mathrm{#1}}}

\newcommand{\eMath}{\mathrm{e}}

\newcommand{\ket}[1]{| {#1}  \rangle }	
\newcommand{\bra}[1]{\langle {#1}  | }	
\newcommand{\iProd}[2]{\ensuremath{\langle \, {#1} | \, {#2} \, \rangle}}
\newcommand{\avg}[1]{\ensuremath{\left\langle {#1} \right\rangle}}

\newcommand{\diff}{\mathrm{d}}
\newcommand{\tDiff}[2]{\dfrac{\diff {#1}}{\diff {#2}}}

\begin{document}

\title{Nonradiative Recombination of Excitons in Carbon Nanotubes\\Mediated by Free Charge Carriers}
\author{J.M. Kinder}
	\email{jesse.kinder@gmail.com}
	\affiliation{Department of Physics, University of Pennsylvania, Philadelphia, Pennsylvania 19104}
\author{E.J. Mele}
	\affiliation{Department of Physics, University of Pennsylvania, Philadelphia, Pennsylvania 19104}
\date{\today}

\begin{abstract}

Free electrons or holes can mediate the nonradiative recombination of excitons in carbon nanotubes. Kinematic constraints arising from the quasi one-dimensional nature of excitons and charge carriers lead to a thermal activation barrier for the process. However, a model calculation suggests that the rate for recombination mediated by a free electron is the same order of magnitude as that of two-exciton recombination. Small amounts of doping may contribute to the short exciton lifetimes and low quantum yields observed in carbon nanotubes.

\end{abstract}


\maketitle

Carbon nanotubes absorb far more light than they emit. Photoluminescence studies of carbon nanotubes typically report quantum yields on the order of 0.01--7\%,\cite{carlson2007fei, jones2005aps, crochet2007qyh, hartschuh2005scn, lefebvre2006pis, russo2006odd, hagen2005edl} which indicates that more than 90\% of the energy absorbed by a carbon nanotube is dissipated by nonradiative processes. In this paper, we identify a nonradiative process that occurs in doped nanotubes: Auger recombination mediated by a free charge carrier.

A photon absorbed by a nanotube can excite an electron from the valence band to the conduction band. The Coulomb interaction allows the excited conduction electron and the empty state in the valence band to bind into a strongly correlated particle-hole pair called an exciton. Excitons created by the absorption of a photon are called ``bright excitons'' since they can couple to the electromagnetic field and recombine by emitting a photon. Carbon nanotubes also support a large number of ``dark excitons'' that cannot recombine by emitting a photon because of conservation laws and selection rules.\cite{dresselhaus2007epc, barros2006sro} In photoluminescence experiments, absorption of light creates a population of bright excitons. Many of these scatter into dark states with lower energy as the exciton population approaches thermal equilibrium.

The quantum yield in a photoluminescence experiment is the ratio of the energy re-radiated by the nanotube sample to the energy absorbed. In quantum dots and dyes, quantum yields often approach 100 percent, so the low yields in carbon nanotubes indicate efficient nonradiative pathways not present in other low-dimensional systems. The processes thought to be responsible fall into two broad classes: exciton transfer in nanotube bundles and nonradiative recombination in individual nanotubes.

Isolated metallic nanotubes have an efficient nonradiative recombination pathway. Because there is no band gap, a particle-hole pair in the lowest band of a metallic nanotube can relax to the ground state through a series of transitions mediated by acoustic phonons. As a result, excitons (created in higher subbands) have short lifetimes and the quantum yield of a metallic nanotube is effectively zero.

This recombination pathway is not available in isolated semiconducting nanotubes. Excited particle-hole pairs will relax to exciton states in the lowest band. However, recombination requires the release an amount of energy equal to the difference between the band gap and the exciton binding energy, typically on the order of a few hundred \unit{meV}. Excitons in semiconducting nanotubes can recombine by emitting a photon, through many-body Auger processes, or by multi-phonon processes. (Multi-phonon decay processes are generally allowed in semiconducting nanotubes but are weak because they require a series of virtual transitions or the simultaneous emission of several phonons.)

When nanotubes are bundled together, excitons migrate from one species to another. Excitons in nanotubes with a large band gap can reduce their energy by hopping to a nearby nanotube with a smaller band gap. Exciton transfer has been observed experimentally in bundles of semiconducting nanotubes.\cite{torrens2006pic} If the bundle also contains metallic nanotubes, all excitons in the bundle can recombine through the efficient nonradiative pathway available in the metallic nanotubes.

Exciton transfer suggests the quantum yield of bundles containing metallic nanotubes will be low. Crochet \emph{et al.}~report that the quantum yield of a suspension of nanotubes increased by two orders of magnitude when bundles were removed, supporting this explanation.\cite{crochet2007qyh} Exciton transfer provides an explanation for the lowest quantum yields reported; however, the largest yields are still only a few percent. The estimated quantum yield for isolated nanotubes suspended in air is about 7 percent,\cite{lefebvre2006pis} suggesting that the intrinsic yield of individual nanotubes is low.

\begin{figure}[hbt]
	\begin{center}
		\centering
	\subfigure[Exciton-Exciton Recombination]{\includegraphics[width=0.4\textwidth]{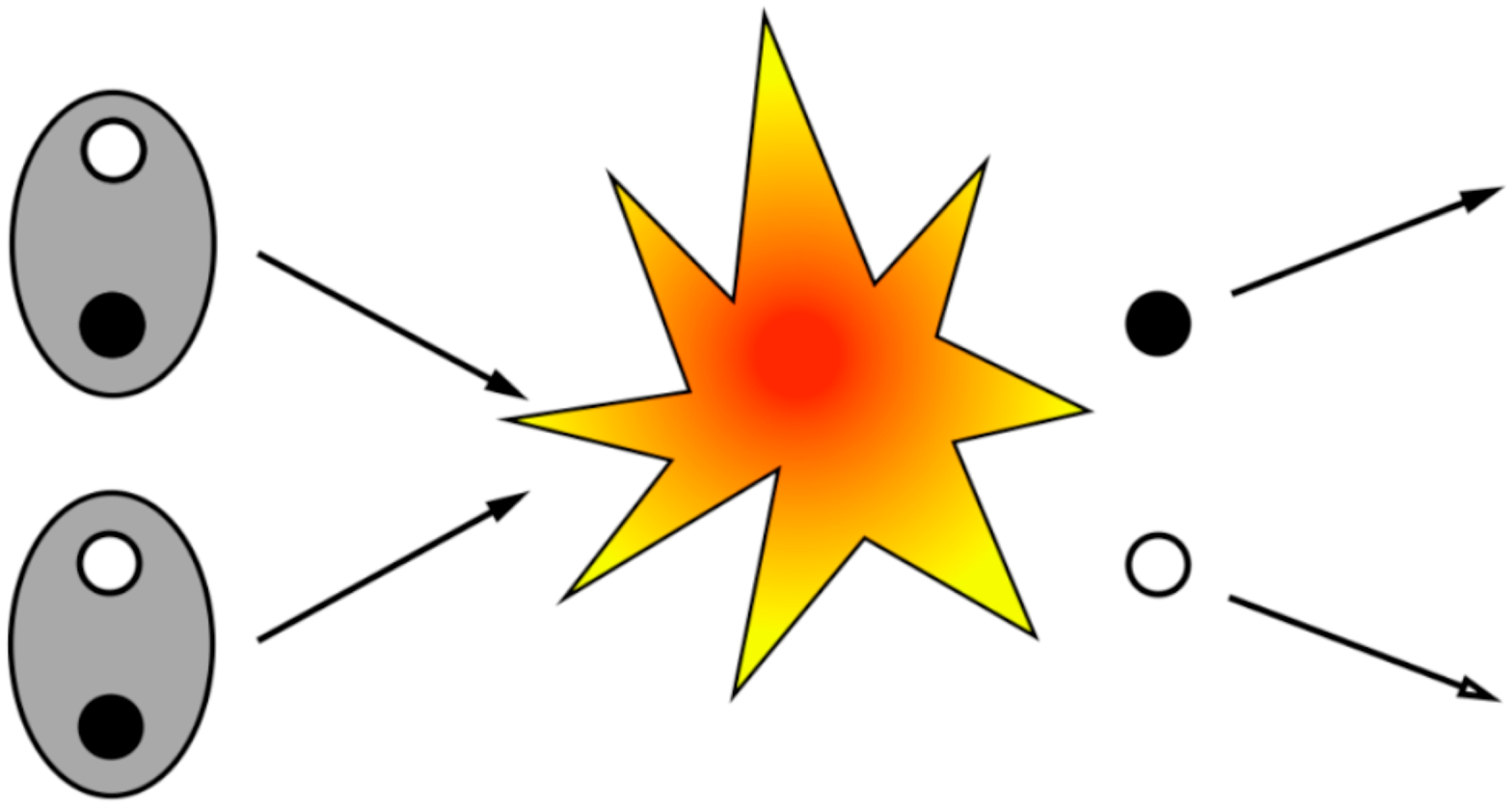}}\qquad \qquad
	\subfigure[Exciton-Electron Recombination]{\includegraphics[width=0.4\textwidth]{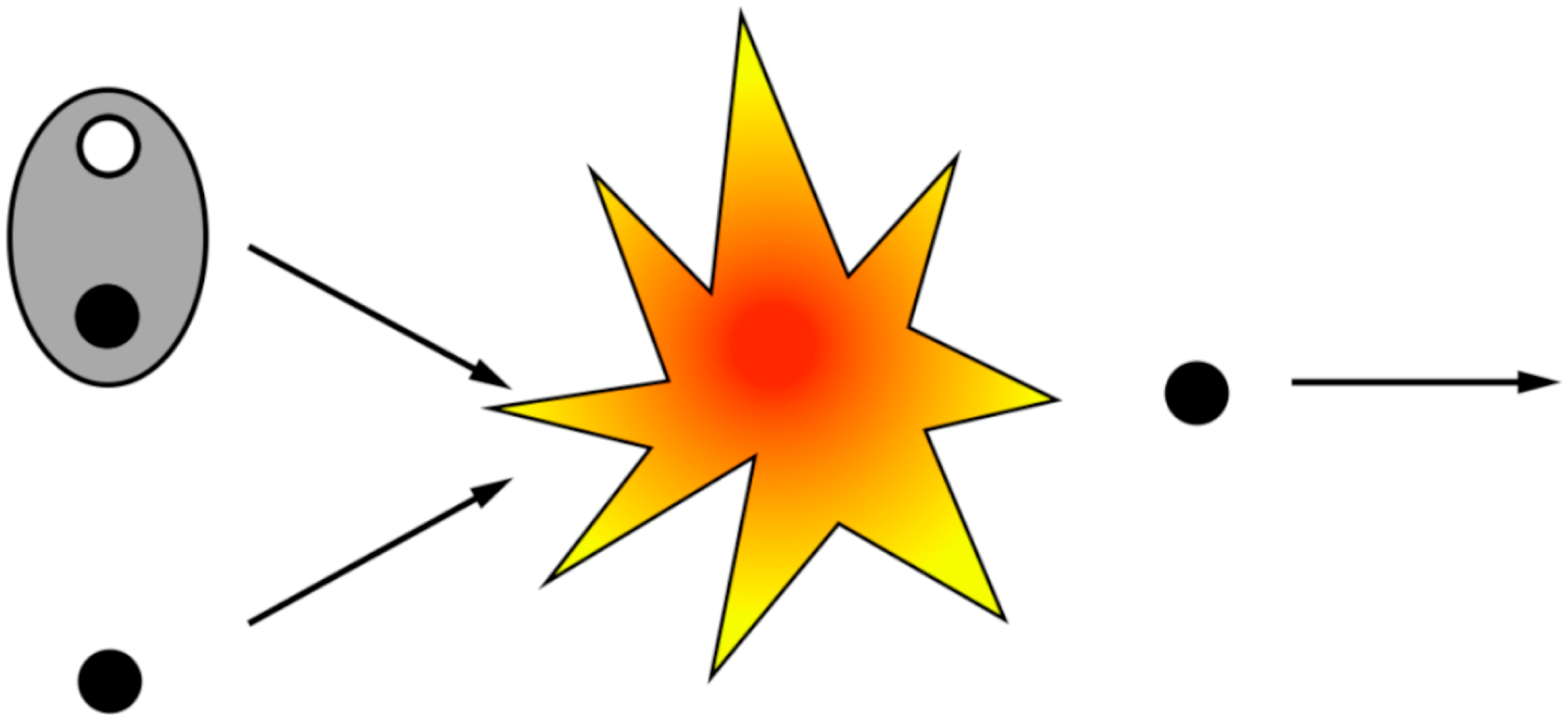}}
		\caption{(Color online) Schematic of two types of nonradiative Auger processes. In exciton-exciton recombination, a particle and a hole recombine and transfer their energy and momentum to the outgoing particle-hole pair. In exciton-electron recombination, all of the energy and momentum are transferred to a single outgoing particle.}
		\label{fig:annihilation}
	\end{center}
\end{figure}

The remainder of this paper describes a class of nonradiative processes in isolated semiconducting carbon nanotubes. We have calculated the decay rate of an exciton population due to an Auger process involving an exciton and a free charge carrier. At room temperature, the rate is comparable to that of two-exciton recombination, giving an exciton lifetime of a few picoseconds. \figRef\ref{fig:annihilation} is a schematic illustration of the two different processes.

The paper is organized as follows. First, we summarize a model developed by Wang, Wu, Hybertsen, and Heinz (WWHH) to describe two-exciton recombination in one-dimensional systems.\cite{wang2006are} Next, we use the model to analyze the nonradiative recombination of an exciton in the presence of a free charge carrier. Finally, we compare the rates of the two processes and discuss the relevance of this decay process to photoluminescence experiments in carbon nanotubes. The details of the rate calculation are included as an appendix.

\section{Exciton Population Dynamics}

Information about the dynamics of the exciton population $N(t)$ comes from spectroscopy, such as time-resolved fluorescence spectroscopy and transient absorption spectroscopy. In order to make a connection with experimental data, a theoretical model of exciton annihilation has to describe $N(t)$, the exciton population after time $t$.  Fermi's Golden Rule translates the amplitude for some quantum mechanical process into a transition rate.\cite{sakurai1994mqm} In this way, one can use quantum mechanics to derive the decay rate of the exciton population:
\begin{equation}
	\tDiff{N}{t} = -\Gamma[N(t)] .
\end{equation}
The decay rate is a function of the population. A simple example is $\Gamma[N(t)] = - \gamma N$, which leads to exponential decay. Multi-exciton processes lead to nonlinear dynamics and power law decay.

Experiments reveal power-law decay at short times ($t \alt 2$ \unit{ps}), followed by exponential decay at longer times ($t \agt 5$ \unit{ps}).\cite{ma2004ucd, wang2004ora, jones2005aps, hagen2005edl, huang2006qba, russo2006odd, valkunas2006eea} This suggests that there are highly efficient multi-exciton decay processes that dominate the dynamics at high exciton densities. At lower densities, single-exciton decay is the only available relaxation pathway.

\subsection{Exciton-Exciton Recombination}

When the exciton density is high, two-body collisions will be frequent. This allows for nonradiative Auger processes of the type $X + X \to p + h$. Two excitons ($X+X$) interact and a particle and a hole recombine. The energy liberated is transferred to the remaining particle and hole ($p+h$). The process is nonradiative since all of the energy and momentum are transferred to the outgoing electrons.

The decay rate is proportional to the square of the exciton population:
\begin{equation}
	\tDiff{N}{t} = - \lambda N^2 ,
\end{equation}
which leads to power law decay. This type of process was first proposed by Ma \emph{et al.}~in 2004 to explain the power law decay in fluorescence and transient absorption measurements on nanotubes.\cite{ma2004ucd}

WWHH developed a theoretical model for a two-exciton Auger recombination process in a one-dimensional system.\cite{wang2006are} The key features of the model are the following:
\begin{itemize}
\item Electrons and holes are described using a two-band model with an allowed optical transition.
\item The Coulomb interaction is replaced by a point-contact interaction.
\item Transition amplitudes are calculated to leading order in \kp{} perturbation theory.
\end{itemize}

\begin{figure}[hbt]
	\begin{center}
	\includegraphics[width=0.48\textwidth]{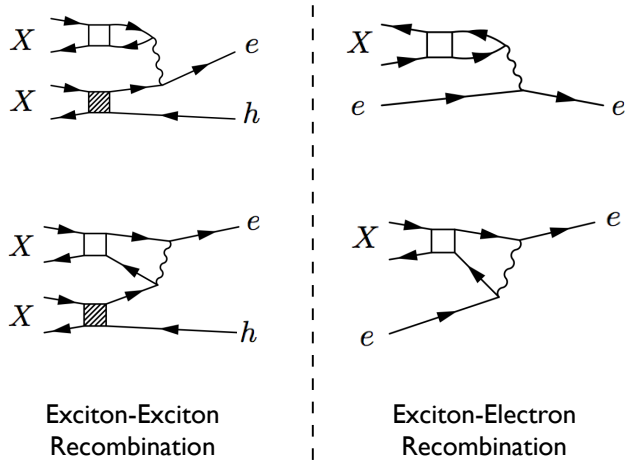}
		\caption{(Color online) Feynman diagrams used to calculate scattering amplitudes in Auger recombination processes. The diagrams on the left were analyzed by Wang, Wu, Hybertsen, and Heinz in Ref.~\onlinecite{wang2006are}. In these nonradiative two-exciton Auger recombination processes, a particle and a hole recombine, transferring their energy and momentum to the scattered particle-hole pair. The diagrams on the right are the subject of this paper. They describe the nonradiative recombination of a single exciton mediated by a free charge carrier.}
		\label{fig:feynmanGraphs}
	\end{center}
\end{figure}

This model is described in more detail in the appendix. Two representative scattering processes from the calculation are shown in \figRef\ref{fig:feynmanGraphs}. Using this model, WWHH calculate a lifetime of $1.7$ \unit{ps} for two excitons in a 1 \unit{\mu m} carbon nanotube,\cite{wang2006are} in reasonable agreement with the experimentally determined decay rate of 3 \unit{ps}.\cite{wang2004ora} This suggests that the model captures the essential features of exciton recombination in carbon nanotubes despite the simplifying approximations.

\subsection{Exciton-Electron Recombination}

Exciton-exciton recombination partially explains the low quantum yields in carbon nanotubes. However, at low exciton densities, two-body interactions are rare and exciton-exciton recombination is irrelevant, but quantum yields are still low. Jones \emph{et al.}~reported a quantum yield of 0.05 percent in a sample with less than one exciton per nanotube.\cite{jones2005aps} This suggests the existence of efficient nonradiative pathways involving only a \emph{single} exciton.

One possibility is an Auger process involving a free charge carrier and an exciton, $X + p \to p$. This is depicted schematically in \figRef\ref{fig:annihilation}.

An ideal semiconducting nanotube is undoped, so a source of free charge carriers is necessary for this process to occur. One possibility is intentionally doping a nanotube with impurities or a gate voltage. Another source of free charge carriers is the environment. Experiments and calculations have shown that oxygen adsorbed onto the surface of a nanotube depletes the valence band, making the nanotube a \emph{p}-type semiconductor.\cite{collins2000eos, kong2000nmw} Other molecules in the environment can also introduce free charge carriers (usually holes) to the nanotube.\cite{jhi2000epo} Additionally, the two-exciton recombination process described above could generate a small number of free charge carriers in an undoped nanotube. It seems likely that a small number of free charge carriers will be available under typical experimental conditions.

\subsubsection*{Kinematic Constraints}

Exciton-electron recombination has a kinematic constraint that leads to a kinetic energy barrier and a temperature-dependent rate.

Consider the two processes in \figRef\ref{fig:annihilation}. Each scattering process must satisfy two constraints: conservation of energy and conservation of momentum (wave vector). In exciton-exciton recombination, there are two outgoing particles. Their wave vectors can be chosen so that the two constraints are satisfied for any initial state. However, in exciton-electron recombination, there is only a single outgoing particle. Its wave vector can be chosen to satisfy one of the constraints, but the other conservation law constrains the \emph{initial} conditions. Only an exciton and an electron whose wave vectors are related in a specific way can participate in the process.

To determine the constraint, we consider a classical model. Electrons have mass $m$; excitons have mass $M$ and internal energy $W$. The internal energy includes any energy not due to the motion of the center of mass. It describes the energy of formation, plus any energy from the dispersion relation of the relative coordinate. $K$ will denote the wave vector of the exciton. $k$ and $q$ will denote the initial and final wave vector of the free particle, respectively.

Conservation of momentum fixes the wave vector of the outgoing particle:
\begin{equation}
	q = k + K .\label{eq:momentum}
\end{equation}
Conservation of energy gives
\begin{equation}
	\dfrac{\hbar^2 q^2}{2m} = \dfrac{\hbar^2 K^2}{2M} + W + \dfrac{\hbar^2 k^2}{2m} . \label{eq:energy}
\end{equation}
Substituting \eqRef(\ref{eq:momentum}) into \eqRef(\ref{eq:energy}) gives the wave vector of the incoming electron as a function of the wave vector of the exciton:
\begin{equation}
	k(K) = \dfrac{m W}{\hbar^2} \cdot \dfrac{1}{K} - \dfrac{1 - m/M}{2} \cdot K .\label{eq:constraint}
\end{equation}

The relation defines a hyperbola in the $k$-$K$ plane. The hyperbola $k(K)$ can have a minimum depending on the ratio of the effective masses of the free carrier and exciton. If the effective mass of the exciton is larger than that of the free charge carrier ($m/M < 1$) then the hyperbola $k(K)$ has no minimum and passes through $k=0$ when the kinetic energy of the exciton is
\begin{equation}
	\dfrac{\hbar^2 K_0{}^2}{2M} = \dfrac{W}{M/m - 1} . \label{eq:kZero}
\end{equation}
If the effective mass of the exciton is \emph{smaller} than that of the free charge carrier ($m/M > 1$), then the hyperbola $k(K)$ passes through a minimum.

As a result, a free electron may be available to mediate the recombination of an exciton depending on two factors: the exciton mass and the doping density.

\textbf{Exciton Mass}: The minimum electron wave vector required for Auger recombination of light excitons is significant in carbon nanotubes. An exciton, composed of a particle and a hole, normally has a larger effective mass than either of its constituent particles. This is the case for dark excitons, which have an effective mass about three times that of an electron or hole: $M \approx 3 m$.\cite{perebeinos2005rle} In contrast, the bright exciton is strongly coupled to the electromagnetic field and has an anomalous dispersion relation that leads to an effective mass smaller than that of a free particle.\cite{perebeinos2005rle, dresselhaus2007epc}

\textbf{Doping Density}: If a carbon nanotube is doped so that there are $n_d$ free charge carriers per unit length, then the Fermi sea will be filled up to $k_F = \pi n_d / 4$. Only electrons with a wave vector less than or equal to $k_F$ are available for scattering.

\begin{figure}[hbt]
	\begin{center}
		\includegraphics[width=0.5\textwidth]{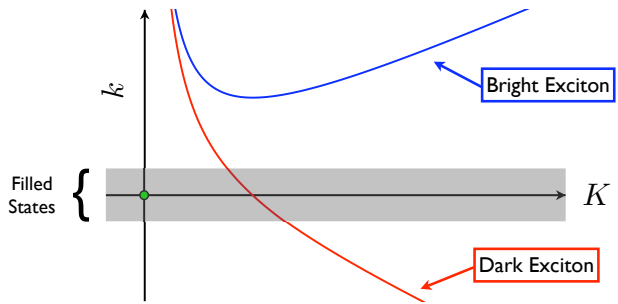}
		\caption{(Color online) The hyperbolas give the allowed electron wave vector $k$ and exciton wave vector $K$ required for Auger recombination. The shaded band indicates the filled Fermi sea of free charge carriers available for scattering. The  effective mass of bright excitons in carbon nanotubes is smaller than the mass of a free carrier, and there is a critical doping density before the process can occur. Dark excitons are more massive than a free carrier, and exciton-electron recombination can occur at infinitesimal doping, where $k_F = 0$.}
		\label{fig:hyperbolas}
	\end{center}
\end{figure}

The interplay between the doping density and the exciton mass is illustrated in \figRef\ref{fig:hyperbolas}. The shaded band represents the filled Fermi sea. The figure shows that electron-assisted recombination of the bright exciton is forbidden unless the doping density exceeds a critical value. In contrast, there is no critical doping density for the decay of dark excitons.

The amount of doping due to the environment is likely to be small. The case of infinitesimal doping --- a single particle at the bottom of the conduction band, with $k_F = 0$ --- will place a lower limit on the rate of exciton-electron recombination. Dark excitons with a kinetic energy given by \eqRef(\ref{eq:kZero}) will be able to recombine nonradiatively.

\subsubsection*{Decay Rate}

Kinematics shows that exciton-electron recombination can satisfy conservation of energy and momentum. A quantum mechanical calculation gives the scattering rate. We have calculated the rate using the model developed by WWHH to describe exciton-exciton recombination.\cite{wang2006are} The two processes relevant to exciton-electron recombination are shown in \figRef\ref{fig:feynmanGraphs}.

In one scattering process, the particle and hole in the exciton recombine and transfer their energy and momentum to the free particle. In the other, the free particle recombines with the hole in the exciton and the energy and momentum are carried off by the particle from the exciton. The second scattering amplitude will have a factor of $(-1)$ relative to the first due to the exchange of fermion operators. Fermi's Golden Rule gives the transition rate, $\Gamma(K_0)$, from the combined scattering amplitude of the two processes in \figRef\ref{fig:feynmanGraphs}. The calculation of $\Gamma(K_0)$ is reproduced in the appendix and may also be found in Ref.~\onlinecite{kinder2008phd}.

$\Gamma(K_0)$ is the decay rate for an exciton with wave vector $K_0$, defined by \eqRef(\ref{eq:kZero}). The decay rate of the exciton population is the product of four terms: the transition rate $\Gamma(K_0)$, the number of free charge carriers ($N_d$), the number of excitons ($N_X$), and the fraction of the exciton population with the correct kinetic energy. The latter is given by the Boltzmann weight. The rate of exciton-electron recombination is
\begin{equation}
	\Gamma_{Xe} = \Gamma(K_0) \cdot N_d \cdot N_X \cdot \eMath^{-E_G/3 k_B T} ,\label{eq:myRate}
\end{equation}
where the Boltzmann weight is calculated using \eqRef(\ref{eq:kZero}) with $M = 3 m$ and $W = 2 E_G / 3$, corresponding to a binding energy of 1/3 the band gap.

\section{Comparison}

\begin{figure}[hbt]
	\begin{center}
	\includegraphics[width=0.5\textwidth]{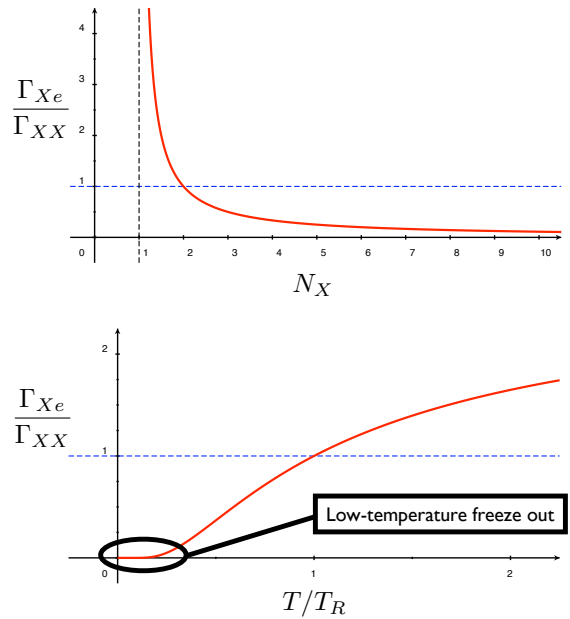}
		\caption{(Color online) Comparison of the rates of exciton-exciton recombination and exciton-electron recombination as a function of exciton population ($N_X$) and temperature as a fraction of room temperature ($T/T_R$). The plots are calculated from \eqRef(\ref{eq:ratio}) for a 1 \unit{\mu m} nanotube with a band gap of 300 \unit{meV}, assuming an exciton radius of 1 \unit{nm} and the presence of a single free electron. The horizontal lines indicate where the rates are equal: two excitons at room temperature.}
		\label{fig:ratePlots}
	\end{center}
\end{figure}

Since $\Gamma_{Xe}$, the population decay rate due to exciton-electron recombination, was calculated within the same model as the exciton-exciton recombination rate ($\Gamma_{XX}$) derived by WWHH in Ref.~\onlinecite{wang2006are}, a direct comparison is possible. Taking the ratio of the two rates eliminates common phenomenological parameters and illustrates the conditions in which one process will be favored over the other. The ratio is
\begin{equation}
	\dfrac{\Gamma_{Xe}}{\Gamma_{XX}} \sim \dfrac{1}{20} \cdot \dfrac{N_d}{N_X - 1} \cdot \dfrac{L}{R_X} \cdot \eMath^{- E_G / 3 k_B T} .\label{eq:ratio}
\end{equation}
$N_d$ is the number of free charge carriers and $N_X$ is the number of excitons in the nanotube. $L$ is the nanotube length and $R_X$ is the exciton radius. $E_G$ is the nanotube band gap, and $T$ is the temperature of the exciton population. \figRef\ref{fig:ratePlots} shows how the ratio varies with the exciton population and the temperature.

\eqRef(\ref{eq:ratio}) is our primary result. It is the product of four terms, each of which will be discussed in turn.

The prefactor of $1/20$ comes from numerical factors arising in the calculation as well as intrinsic properties of the nanotube, such as the band gap and exciton binding energy. This term favors exciton-exciton recombination. One contribution to this factor is the number of processes that contribute to the total scattering amplitude. In two-exciton recombination, there are four possible ways for a particle and hole to recombine. In addition, there are two choices for which outgoing particle is involved in the interaction, giving a total of 8 different scattering processes. For exciton-electron recombination, there are only two possibilities. (See \figRef\ref{fig:feynmanGraphs}.) 

The factor of $N_d/(N_X - 1)$ favors the two-exciton process at high exciton densities, but goes to infinity when the number of excitons falls to 1 (see \figRef\ref{fig:ratePlots}). This divergence reflects the fact that a single exciton cannot decay by a two-exciton process.

The third factor, $L/R_X$, is the ratio of the system size --- the length of the nanotube --- to the exciton radius, which is on the order of the nanotube radius. It strongly favors exciton-electron recombination. The physical meaning of this term lies in the finite extent of the exciton wave function. For two excitons to recombine, their wave functions must overlap. The exciton-electron process involves a Bloch electron, which is delocalized throughout the entire system. The electron wave function \emph{always} overlaps the exciton wave function.

The Boltzmann weight $\exp ( - E_G / 3 k_B T )$ is due to the kinetic energy barrier of exciton-electron recombination. Excitons are created with no net momentum, and all of their kinetic energy comes from thermal fluctuations. At low temperatures, exciton-electron recombination will be frozen out, but at high temperatures, the Boltzmann weight will be irrelevant (see \figRef\ref{fig:ratePlots}). An important caveat is that the temperature in this expression is the effective temperature of the exciton population. Although most experiments are performed at room temperature, the lasers used to generate excitons could cause significant heating of the nanotube, resulting in a much larger effective temperature.

Despite the numerical prefactor and kinetic energy barrier, the rate of exciton-electron recombination is comparable to two-exciton recombination. Consider two excitons ($N_X = 2$) in an isolated nanotube with a length of 1 \unit{\mu m} and a band gap of 300 \unit{meV}. At room temperature ($k_B T \approx 25$ \unit{meV}) the two rates in \eqRef(\ref{eq:ratio}) are roughly equal if there is just \emph{one} free electron in the nanotube. This suggests that Auger recombination of dark excitons mediated by a free charge carrier leads to an exciton lifetime on the order of a few \unit{ps}. Exciton-electron recombination could provide an efficient nonradiative decay mechanism for dark excitons in carbon nanotubes.

\section{Experimental Signature}

\eqRef(\ref{eq:ratio}) suggests two experimental signatures of Auger recombination mediated by free charge carriers.

First, the decay rate is proportional to the doping density and will increase linearly with the number of free charge carriers at low doping densities. (At higher doping densities, the approximation of infinitesimal doping breaks down and a modified calculation of the rate is necessary.) The number of free charge carriers could be controlled with a gate voltage for nanotubes deposited on a substrate. Another possibility is intentionally doping nanotubes with specific surfactants or solvents. For nanotubes in solution, the quantum yield varies with the pH.\cite{oconnell2002bgf} Exciton-electron recombination could be relevant to this effect.

The second experimental signature of Auger recombination mediated by free charge carriers is a freeze out at low temperatures. For $k_B T \ll E_G$, no excitons will acquire the necessary kinetic energy to decay by this process. The decay rate should decrease with decreasing temperature, as should the quantum yield. A plot of the logarithm of the decay rate versus the inverse temperature should exhibit linear scaling.

These effects could be extracted from the exponential tails of fluorescence and transient absorption measurements. At short times, exciton-exciton recombination is the dominant relaxation pathway and the effects of single-exciton processes would be difficult to extract from the data.

\section{Summary}

A model calculation suggests that Auger recombination mediated by free charge carriers in carbon nanotubes could provide an efficient nonradiative decay channel for dark excitons. Nanotubes may be doped --- intentionally or unintentionally --- and this would provide a population of free charge carriers that could participate in Auger recombination processes. As a result, the decay rate of a dilute exciton population could be highly sensitive to doping and the temperature of the sample.

Recombination mediated by free charge carries is not the only candidate for an efficient nonradiative decay process. Perebeinos \emph{et al.}~have analyzed multi-phonon decays and phonon-assisted decays in doped carbon nanotubes as well as the effects of exciton localization.\cite{perebeinos2008pae} Experimental measurements of the relation between exciton lifetimes, the doping density, and temperature should give more insight into the mechanism responsible for the low quantum yields in carbon nanotubes.

More detailed calculations might provide new insight as well. The model presented here could be extended in several ways. One is to include the effects of two degenerate valleys in the nanotube band structure. This is unlikely to have a significant effect on the result. The symmetry of a dark exciton, which forbids its own radiative recombination, would only have a small effect on the calculated decay rate. The dominant contribution to the decay rate comes from the scattering process in which a free electron recombines with the hole in the exciton. The antisymmetry of the exciton wave function does not have a significant effect on the probability of finding a \emph{free} electron and the bound hole in the same place.

It seems likely that free charge carriers in carbon nanotubes can have a significant effect on exciton population dynamics, but further study is necessary to definitively establish their role in the low quantum yields reported in photoluminescence experiments.

This work was supported by the Department of Energy under grant DE-FG02-84ER45118.



\appendix

\section{Details of Rate Calculation}

The following calculation is reproduced from Ref.~\onlinecite{kinder2008phd}. It is based on a model developed by Wang, Wu, Hybertsen, and Heinz in Ref.~\onlinecite{wang2006are}.

Consider a two-band model with a point interaction in a one-dimensional system of length L. Particles and holes are described by Bloch states:
\begin{equation}
	\ket{\psi_p(k)} = \eMath^{i k z} \ket{u_c(k)} \qquad \ket{\psi_h(k)} = \eMath^{-i k z} \ket{u_v(-k)}.
\end{equation}
Each band is assumed to be parabolic with effective mass $m$:
\begin{equation}
	\varepsilon_{\pm}(k) = \pm \left( \dfrac{E_G}{2} + \dfrac{ \hbar^2 k^2 }{ 2 m } \right).
\end{equation}

The Coulomb interaction is replaced by a short-range potential:
\begin{equation}
	V(z) \longrightarrow - U \, \delta(z) .
\end{equation}
The Fourier transform of the interaction potential is $V(q) = U/L$.

This potential allows for excitons. It creates bound particle-hole pairs inside the band gap, with a binding energy of
\begin{equation}
	E_B = \dfrac{m U^2}{\hbar^2} .\label{eq:xBind}
\end{equation}
The bound states are exponentially localized, with a spatial extent of
\begin{equation}
	\dfrac{1}{\kappa} = \dfrac{2 \hbar^2}{m U} .
\end{equation}

The exciton wave function is the product of functions describing the center of mass and the separation of the particle-hole pair:
\begin{equation}
	\Psi(z_p,z_h) = \Phi(Z) \cdot \phi(z),
\end{equation}
where $Z = (z_p + z_h)/2$ and $z = z_p - z_h$. The center of mass is described by a plane wave with wave vector $K = k_e + k_h$, and the relative coordinate is exponentially localized:
\begin{equation}
	\phi(z) = \sqrt{\dfrac{\kappa}{2}} \, \eMath^{-\kappa |z|} .
\end{equation}
The Fourier transform of $\phi(z)$ is
\begin{equation}
	\phi(q) = \dfrac{1}{\sqrt{\kappa L}} \cdot \dfrac{2 \kappa^2}{\kappa^2 + q^2} ,
\end{equation}
where $q = (k_p - k_h)/2$ is the relative wave vector.

The exciton state is
\begin{align}
	\ket{\Psi(K)} &= \sum_{k_p,k_h} \, \Phi_{K}(k_p,k_h) \, \ket{u_c(k_p)} \otimes \ket{u_v(-k_h)} \\
		&= \sum_{k_p,k_h} \, \delta_{K, k_p + k_h } \, \dfrac{1}{\sqrt{\kappa L}} \cdot \dfrac{2 \kappa^2}{\kappa^2 + q^2} \, \ket{u_c(k_p)} \otimes \ket{u_v(-k_h)}. \label{eq:xWave}
\end{align}

Matrix elements of the Coulomb interaction are evaluated using $\mathbf{k \cdot p}$ perturbation theory:
\begin{align}
	\ket{u_{c}(k)} &\approx \ket{u_{c}(k_{0})} + \dfrac{\hbar}{m_e} \dfrac{ \bra{u_{v}(k_{0})} (k - k_0) \cdot p \ket{u_{c}(k_{0})} }{ \varepsilon_{c}(k_{0}) - \varepsilon_{v}(k_{0}) } \, \ket{u_{v}(k_{0})}, \label{eq:pBloch} \\
	\ket{u_{v}(k)} &\approx \ket{u_{v}(k_{0})} + \dfrac{\hbar}{m_e} \dfrac{ \bra{u_{c}(k_{0})} (k - k_0) \cdot p \ket{u_{v}(k_{0})} }{ \varepsilon_{v}(k_{0}) - \varepsilon_{c}(k_{0}) } \, \ket{u_{c}(k_{0})}. \label{eq:hBloch}
\end{align}
$m_e$ is the free electron mass, not the effective mass of the charge carriers.

Using this approximation, all of the inner products in the calculation are 1 or can be expressed in terms of the dipole transition amplitude,
\begin{equation}
	\bra{u_{c}(k_{0})} (k - k_{0}) \cdot p \ket{u_{v}(k)} = (k - k_{0}) \cdot \langle p \rangle_{cv} .
\end{equation}

For electrons in the same band,
\begin{equation}
	\iProd{u_{c}(k_{0})}{u_{c}(k_{0}+k)} = \iProd{u_{v}(k_{0})}{u_{v}(k_{0}+k)} \approx 1 .\label{eq:ccvvAmp}
\end{equation}
For those in opposite bands,
\begin{equation}
	\iProd{u_{v}(-k_{h})}{u_{c}(k_{p})} \approx \dfrac{\hbar}{m_e} \dfrac{ k_{p} + k_{h} }{ E_G } \langle p \rangle_{cv}.\label{eq:cvvcAmp}
\end{equation}
To obtain this expression, the energy difference between the conduction and valence bands is approximated by the band gap. This will place an upper limit on the amplitude.

These rules can be used to evaluate the total amplitude $\mathcal{M}_{io}$ for a scattering process, where $i$ and $o$ indicate the initial and final states. Fermi's Golden Rule gives the transition rate due to the scattering process:
\begin{equation}
	\Gamma = \dfrac{2 \pi}{\hbar} \sum_o | \mathcal{M}_{io} |^2 \cdot n_i \cdot [ 1 - n_o ] \cdot \delta( \varepsilon_i - \varepsilon_o ).
\end{equation}
$n_i$ and $n_o$ are the occupation probabilities of the initial and final states.

The two exciton-electron scattering processes in \figRef\ref{fig:feynmanGraphs} give the decay rate for Auger recombination mediated by a free charge carrier. In the calculation below, $k$ is the wave vector of the outgoing particle, and $K$ is the wave vector of the exciton. To satisfy conservation of momentum, the wave vector of the incoming free particle must be $k - K$.

In one process, the particle and hole in the exciton recombine and transfer their energy and momentum to the free particle. The amplitude is given by
\begin{align}
	\mathcal{A} &= \sum_{k_{p},k_{h}} V(K) \, \Phi_K(k_p,k_h) \, \iProd{u_v(-k_h)}{u_c(k_p)} \iProd{u_c(k)}{u_c(k-K)} \\
		&= \dfrac{\hbar \avg{p}_{cv} }{m_e E_G} \cdot \dfrac{U}{L} \cdot \sum_{q} \dfrac{1}{\sqrt{\kappa L}} \dfrac{2 \kappa^2}{\kappa^2 + (q - K/2)^2} \cdot K \\
		&\approx \dfrac{\hbar \avg{p}_{cv} }{m_e E_G} \cdot \dfrac{U}{L} \cdot \sqrt{\kappa L} \cdot K.
\end{align}
The \deltaFunction{} in \eqRef(\ref{eq:xWave}) collapses one of the sums over $k_p$ or $k_h$. The remaining sum may be approximated by an integral over $q$ to obtain the final expression.

In the other process, the free particle recombines with the hole in the exciton and the energy and momentum are carried off by the particle from the exciton. The amplitude may be evaluated in a manner similar to $\mathcal{A}$:
\begin{align}
	\mathcal{B} &= \sum_{k_{p},k_{h}} V(k-k_p) \, \Phi_K(k_p,k_h) \, \iProd{u_c(k)}{u_c(k_p)} \iProd{u_v(-k_h)}{u_c(k-K)} \\
		&\approx \dfrac{\hbar \avg{p}_{cv} }{m_e E_G} \cdot \dfrac{U}{L} \cdot (k - K/2) \cdot \sqrt{\kappa L}.
\end{align}

$\mathcal{B}$ will have a factor of $(-1)$ relative to $\mathcal{A}$ due to the exchange of fermion operators. The total amplitude for the process is
\begin{equation}
	\mathcal{M}(k) = \mathcal{A} - \mathcal{B} \approx - \dfrac{\hbar \avg{p}_{cv} }{m_e E_G} \cdot \dfrac{U}{L} \cdot \sqrt{\kappa L} \cdot (k - 3K/2).
\end{equation}
At infinitesimal doping, the wave vector of the incoming free particle is 0. Conservation of momentum requires that $k = K$, so
\begin{equation}
	\mathcal{M}(K) = \dfrac{\hbar \avg{p}_{cv} }{m_e E_G} \cdot \dfrac{U}{L} \cdot \sqrt{\kappa L} \cdot \dfrac{K}{2}.
\label{eq:decayAmp}
\end{equation}

Fermi's Golden Rule gives the decay rate:
\begin{equation}
	\Gamma(K) = \dfrac{2 \pi}{\hbar} \sum_K | \mathcal{M}(K) |^2 \cdot n(0) \cdot [ 1 - n(K) ] \cdot \delta[ E_X(K) + \varepsilon_p(0) - \varepsilon_p(K) ]. \label{eq:fermi}
\end{equation}

The \deltaFunction{} enforces conservation of energy and is equivalent to \eqRef(\ref{eq:kZero}). Using $M = 3 m$,
\begin{equation}
	\delta[ E_X(K) + \varepsilon_p(0) - \varepsilon_p(K) ] \approx  \dfrac{M}{\hbar^2 K_0} \cdot \dfrac{L}{2\pi} \delta_{K,K_0}, \label{eq:fermiWeight}
\end{equation}
where $\hbar K_0 = \sqrt{3 m W}$.

From \eqRef(\ref{eq:decayAmp}),
\begin{equation}
	| \mathcal{M}(K) |^2 = \dfrac{ \hbar^2 K^2 }{ 4 m_{e}^{2} } \dfrac{ |\langle p \rangle_{cv} |^2 }{ E_G ^2 } \dfrac{ \kappa U^2 }{ L } .
\end{equation}
\eqRef(\ref{eq:xBind}) gives
\begin{equation}
	\dfrac{ \kappa U^2 }{ L } = \dfrac{ E_{B}^2 }{ \kappa L }.
\end{equation}
Therefore,
\begin{equation}
	|\mathcal{M}(K)|^2 = \dfrac{ \hbar^2 K^2 }{ 4 m_e } \cdot \dfrac{ E_{B}^2 }{ E_{G}^{2} } \cdot \dfrac{ 1 }{ \kappa L} \cdot \dfrac{ |\langle p \rangle_{cv} |^2 }{ m_e }. \label{eq:fermiAmp}
\end{equation}

With \eqRef(\ref{eq:fermiWeight}) and (\ref{eq:fermiAmp}), the transition rate is
\begin{equation}
	\Gamma(K_0) = \dfrac{1}{4} \left( \dfrac{E_B}{E_G} \right)^2 \cdot \dfrac{K_0}{\kappa} \cdot \dfrac{3 m}{m_e} \cdot \dfrac{ |\langle p \rangle_{cv} |^2 }{ \hbar m_e } .
\end{equation}

This is the decay rate for an exciton with wave vector $K_0$. The decay rate of the exciton population is the product of four terms: the rate $\Gamma(K_0)$, the number of free charge carriers ($N_d$), the number of excitons ($N_X$), and the fraction of the exciton population with kinetic energy $W/2 = E_G/3$. The population decay rate due to exciton-electron recombination is
\begin{align}
	\Gamma_{Xe} &= N_d \cdot N_X \cdot \eMath^{-E_G/3 k_B T} \cdot \Gamma(K_0) \\
		&= N_d \cdot N_X \cdot \eMath^{- E_G/ 3 k_B T} \cdot \dfrac{1}{4} \left( \dfrac{E_B}{E_G} \right)^2 \cdot \dfrac{K_0}{\kappa} \cdot \dfrac{3 m}{m_e} \cdot \dfrac{ |\langle p \rangle_{cv} |^2 }{ \hbar m_e } .\label{eq:myRate}
\end{align}

In Ref.~\onlinecite{wang2006are}, WWHH calculated the decay rate of the exciton population due to exciton-exciton recombination to be
\begin{equation}
	\Gamma_{X-X} \approx 128 \cdot N_X \cdot (N_X - 1) \cdot \dfrac{ m }{ 2 m_e } \cdot \left( \dfrac{ E_B }{ E_G } \right)^3 \cdot \dfrac{ 1 }{ k_0 L } \cdot \left(\dfrac{|\langle p \rangle_{cv}|^2}{\hbar m_e} \right) ,\label{eq:them}
\end{equation}
where $N_X$ is the number of excitons in the system and $k_0$ is a wave vector determined by conservation of energy:
\begin{equation}
	\dfrac{\hbar^2 k_0 {}^2 }{m} = E_G - 2 E_B .
\end{equation}
Corrections are of order in $(E_B/E_G)^2$.

Taking the ratio of \eqRef(\ref{eq:myRate}) and \eqRef(\ref{eq:them}) leads to \eqRef(\ref{eq:ratio}).

\end{document}